\documentclass[oneside,onecolumn,12pt,a4paper]{article}
\usepackage{graphicx}
\usepackage{dsfont}
\usepackage{natbib}
\usepackage{setspace}
\usepackage{sectsty}
\usepackage{tikz}
\usepackage{amsmath}
\usepackage{hyperref}
\usepackage{color}
\definecolor{darkblue}{rgb}{0.0,0.0,0.3}
\hypersetup{colorlinks,breaklinks,linkcolor=darkblue,urlcolor=darkblue,anchorcolor=darkblue,citecolor=darkblue}

\def\citepos#1{\citeauthor{#1}'s (\citeyear{#1})}

\usepackage{ifthen}
\def\eprinttmp@#1arXiv:#2 [#3]#4@{\ifthenelse{\equal{#3}{}}{\href{http://arxiv.org/abs/#1}{arXiv:#1}}{\href{http://arxiv.org/abs/#2}{arXiv:#2 [#3]}}}
\newcommand{\eprint}[1]{\eprinttmp@#1arXiv: []@}
\newcommand{\doi}[1]{\href{http://dx.doi.org/#1}{doi:#1}}

\begin{document}

\sectionfont{\normalfont\normalsize\bfseries}

\subsectionfont{\normalfont\small\bfseries}

\renewcommand{\theenumi}{(\roman{enumi})}
\renewcommand{\labelenumi}{\theenumi}

\setstretch{1.4}

\title{Causal Symmetry and the Transactional Interpretation}
 \author{\normalsize{Peter W. Evans}}
 \date{14 April, 2012}
 \maketitle

\begin{abstract}
\citepos{Cramer} transactional interpretation of quantum mechanics posits retrocausal influences in quantum processes in an attempt to alleviate some of the interpretational difficulties of the Copenhagen interpretation. In response to Cramer's theory, \citet{Maudlin} has levelled a significant objection against \emph{any} retrocausal model of quantum mechanics. I present here an examination of the transactional interpretation of quantum mechanics and an analysis of Maudlin's critique. I claim that, although Maudlin correctly isolates the weaknesses of Cramer's theory, his justification for this weakness is off the mark. The cardinal vice of the transactional interpretation is its failure to provide a sufficient causal structure to constrain uniquely the behaviour of quantum systems and I contend that this is due to a lack of \emph{causal symmetry} in the theory. In contrast, Maudlin attributes this shortcoming to retrocausality itself and emphasises an apparently fundamental incongruence between retrocausality and his own metaphysical picture of reality. I conclude by arguing that the problematic aspect of this incongruence is Maudlin's assumptions about what is appropriate for such a metaphysical picture.
\end{abstract}

\section{Introduction}

The merit of positing retrocausal influences in quantum mechanics is relatively well known: explicit violation of the assumption of independence in the derivation of Bell's Inequality resurrects the local hidden variables program of quantum interpretations. Positing retrocausality in nature, however, is a somewhat unpopular proposal. One of the most significant obstacles for retrocausal approaches to quantum mechanics is the objection levelled at \citepos{Cramer} transactional interpretation of quantum mechanics by \citet{Maudlin}, who claims that his objection poses a problem for ``any theory in which both backwards and forwards influences conspire to shape events''. This paper is an examination of Maudlin's objection to retrocausality.

The examination proceeds as follows. I begin in \S\ref{sec:wheeler_feynman} with an introduction to \citepos{WheelerFeynman} attempted time symmetric formulation of classical electrodynamics, from which the transactional interpretation of quantum mechanics originates. I then introduce in \S\ref{sec:transaction} Cramer's extension of the Wheeler-Feynman formalism to a retrocausal transaction mechanism for modelling quantum processes. \S\ref{sec:interpretation} sets out the details of the transactional interpretation and I briefly mention there some of the advantages Cramer's theory has over the Copenhagen interpretation of quantum mechanics: most notably that the retrocausal structure allows a `zigzag' causal explanation of the nonlocality associated with Bell-type quantum systems. In \S\ref{sec:maudlin} I set out the details of Maudlin's inventive thought experiment that constitutes his objection to Cramer's theory. I examine in \S\ref{sec:defend} some replies that have been made in response to Maudlin's objection defending the transactional interpretation.

In \S\ref{sec:fourdim} I offer my own analysis of Maudlin's experiment according to the transactional interpretation with a view to showing that, despite the putative defences considered, there is still a problem to be overcome. What is lacking in Cramer's theory is a causal structure that can constrain uniquely the behaviour of a quantum system and this is exactly the problem that Maudlin's experiment emphasises. I diverge from Maudlin, however, in the justification for why the transactional interpretation suffers this shortcoming. I claim in \S\ref{sec:symmetry} that it is the failure of the transactional interpretation to ensure causal symmetry that is impeding such unique determination of behaviour. In contrast, Maudlin attributes this shortcoming to retrocausality itself and emphasises an apparently fundamental incongruence between retrocausality and his own ``metaphysical picture of the past generating the future''. I present an argument that it is Maudlin's assumption about the appropriateness of this metaphysical picture that is problematic here, and not retrocausality.

\section{The Wheeler-Feynman absorber theory of radiation}
\label{sec:wheeler_feynman}

Our narrative begins with a problem of classical electrodynamics: an accelerating electron emits electromagnetic radiation, and through this process the acceleration of the electron is damped. Various attempts were initially made to account for this phenomenon in terms of the classical theory of electrodynamics but largely these lacked either empirical adequacy or a coherent physical interpretation. \citet{WheelerFeynman} set out to remedy this situation by reinterpreting \citepos{Dirac} theory of radiating electrons. I will make no attempt here to give an analysis of this problem, nor of its ensuing evolution. What is important for our purposes is the nature of the interpretation that Wheeler and Feynman proffer as a resolution, for it is this interpretation that is the motivation for the transactional interpretation of quantum mechanics.

The core of Wheeler and Feynman's absorber theory of radiation is a suggestion that the process of electromagnetic radiation should be thought of as an interaction between a source and an absorber rather than as an independent elementary process.\footnote{Such an idea was suggested, for instance, by \citet{Tetrode} and also by \citet{LewisGN}:
\begin{quote}
  [A]n atom never emits light except to another atom, and\ldots it is as absurd to think of light emitted by one atom regardless of the existence of a receiving atom as it would be to think of an atom absorbing light without the existence of light to be absorbed. I propose to eliminate the idea of mere emission of light and substitute the idea of transmission, or a process of exchange of energy between two definite atoms or molecules.~\citep[p.~24]{LewisGN}
\end{quote}
} Wheeler and Feynman imagine an accelerated point charge located within an absorbing system and consider the nature of the electromagnetic field associated with the acceleration. An electromagnetic disturbance initially travels outwards from the source and perturbs each particle of the absorber. The particles of the absorber then generate together a subsequent field. According to the Wheeler-Feynman view, this new field is comprised of half the sum of the retarded (forwards-in-time) and advanced (backwards-in-time) solutions to Maxwell's equations. The sum of the advanced effects of all the particles of the absorber then yields an advanced incoming field that is present at the source simultaneous with the moment of emission. The claim is that this advanced field exerts a finite force on the source which has exactly the required magnitude and direction to account for the observed energy transferred from source to absorber; this is Dirac's radiative damping field. In addition, when this advanced field is combined with the equivalent half-retarded, half-advanced field of the source, the total observed disturbance is the full retarded field known from experience to be emitted by accelerated point charges.

The crucial point to note about the Wheeler-Feynman scheme is that due to the \emph{advanced} field of the absorber, the radiative damping field is present at the source at exactly the time of the initial acceleration. Quite simply, if a retarded electromagnetic disturbance propagates for a time $t$ before meeting the absorber then the absorber will be a distance $ct$ from the source. The advanced field propagates with the same speed $c$ across the same distance and thus will arrive at the source exactly time $t$ \emph{before} the absorber field is generated, i.e. at the time of the initial acceleration. If we think of this four dimensionally (in block universe terms) it is clear to see that the advanced field does not simply propagate the same distance as the source field, it propagates across the very same spacetime as the initial disturbance.\footnote{According to \citet{Price91}, the fact that the retarded and advanced waves cross the same spacetime indicates that they are in fact one and the same electromagnetic disturbance.} It is this idea of time symmetric radiation that is at the core of the transactional interpretation of quantum mechanics.

\section{The quantum handshake}
\label{sec:transaction}

\citepos{Cramer} transactional interpretation is a retrocausal model of quantum mechanics that extends the Wheeler-Feynman formalism beyond electrodynamics. Cramer suggests that the description of the emission and absorption of electromagnetic radiation in the Wheeler-Feynman scheme can be adopted to describe the microscopic exchange of a single quantum of energy, momentum, \emph{etc.}, between and within quantum systems. This time symmetric interpretation of the quantum mechanical formalism not only provides a Lorentz covariant explanation of the nonlocal behaviour displayed by entangled quantum systems but also constitutes an attempt to alleviate some of the interpretational problems of the Copenhagen interpretation in general. Before we address how this is achieved, let us consider the transaction mechanism at the core of the transactional interpretation.

Imagine a quantum emitter such as a vibrating electron or atom in an excited state. According to Cramer, when a single quantum is to be emitted (a photon, in these cases) the source produces a radiative field. Analogously to the Wheeler-Feynman description, this field propagates outwards in all directions of four dimensional spacetime, i.e. in all three spatial dimensions and both forwards (retarded field) and backwards (advanced field) in the temporal dimension. When this field encounters an absorber, a new field is generated that likewise propagates in all directions of four dimensional spacetime. The retarded field produced by the absorber exactly cancels the incident retarded field produced by the emitter for all times after the absorption of the photon. The advanced field produced by the absorber propagates backwards in time across the same spacetime interval as the incident wave to be present at the emitter at the instant of emission. The advanced field produced by the absorber exactly cancels the advanced field produced by the emitter and thus there is neither a net field present after the time of absorption nor before the initial emission; only between the emitter and the absorber is there a radiative field.

Cramer describes the field that travels from the source to the absorber as an ``offer'' wave and the field that returns from the absorber to the emitter as a ``confirmation'' wave. The transaction is completed with a ``handshake'': the offer and confirmation waves combine to form a four dimensional standing wave between emitter and absorber. The conditions at the emitter and absorber at the time of emission and absorption respectively are the boundary conditions that determine whether or not a transaction can take place and, if so, the probability of that transaction occurring. The amplitude of the confirmation wave which is produced by the absorber is proportional to the local amplitude of the incident wave that stimulated it and this, in turn, is dependent on the attenuation it received as it propagated from the source. It is the boundary conditions at both ends of the transaction that define when a transaction can be completed. A cycle of offer and confirmation waves ``repeats until the response of the emitter and absorber is sufficient to satisfy all of the quantum boundary conditions\ldots at which point the transaction is completed'' \citeyearpar[p.~662]{Cramer}. Many confirmation waves from potential absorbers may converge on the emitter at the time of emission but the quantum boundary conditions can usually only permit a single transaction to form. Any observer who witnesses this process would perceive only the completed transaction, which would be interpreted as the passage of a particle (e.g. a photon) between emitter and absorber.

There are in fact two complementary descriptions of the transaction process lurking side by side here: on the one hand there is a description of the physical process, consisting of the passage of a particle between emitter and absorber, that a temporally bound experimenter would observe; and on the other hand there is a description of a dynamical process of offer and confirmation waves that is instrumental in establishing the transaction. This latter process clearly cannot occur in an ordinary time sequence, not least because our temporally bound observer by construction cannot detect any offer or confirmation waves. Cramer suggests that the `dynamical process' be understood as occurring in a ``pseudotime'' sequence:
\begin{quote}
  \small
  \singlespacing
  The account of an emitter-absorber transaction presented here employs the semantic device of describing a process extending across a lightlike or a timelike interval of space-time as if it occurred in a time sequence external to the process. The reader is reminded that this is only a pedagogical convention for the purposes of description. The process is atemporal and the only observables come from the superposition of all ``steps'' to form the final transaction. \citeyearpar[p.~661,~fn.~14]{Cramer}
\end{quote}
These steps are of course the cyclically repeated exchange of offer and confirmation waves which continue ``until the net exchange of energy and other conserved quantities satisfies the quantum boundary conditions of the system'' \citep[p.~662]{Cramer}. There is a strong sense here that any process described as occurring in pseudotime is \emph{not a process at all} but, as Cramer reminds, merely a ``pedagogical convention for the purposes of description''. The role that pseudotime plays in Cramer's theory will be of major concern for us in this analysis and we will see in \S\ref{sec:fourdim} that the ontological status of Cramer's posited pseudotemporal sequence is far from transparent. For now, however, let us consider how this transaction mechanism underpins Cramer's transactional interpretation of quantum mechanics.

\section{The transactional interpretation}
\label{sec:interpretation}

Cramer utilises the principled framework of the Copenhagen interpretation to characterise his transactional interpretation. Recall that the Copenhagen interpretation can be characterised in terms of a clutch of core principles, including Heisenberg's indeterminacy relation, the Born rule, Bohr's principle of complementarity and the epistemic reading of the wavefunction. The purpose of these principled elements is to provide a physical picture of quantum systems given the formalism of quantum mechanics; Cramer likewise constructs the transactional interpretation from principles to serve this end.

To begin with, the statistical interpretation of the formalism embodied in the Born rule remains unchanged from the Copenhagen interpretation. This is a consequence of the fact that during the transaction process the confirmation wave traverses the very same spacetime as the offer wave, only in reverse: the amplitude of the advanced component of the confirmation wave arriving back at the emitter is proportional to the time reverse (or complex conjugate) of the amplitude of the initial offer wave evaluated at the absorber. Thus, the total amplitude of the confirmation wave is just the absolute square of the initial offer wave (evaluated at the absorber), which yields the Born rule. Since the Born rule arises as a product of the transaction mechanism, there is no special significance attached to the role of the observer in the act of measurement. The `collapse of the wave function' is interpreted as the completion of the transaction. Thus both the indeterminacy relation and the principle of complementarity are no longer fundamentally related to the process of observation but rather dissolve into a single feature of the transaction mechanism: in satisfying the boundary conditions, the transaction can project out and localise only one of a pair of conjugate variables.

According to Cramer, the biggest bifurcation between the Copenhagen and transactional interpretations is centred around the physical significance of the wavefunction. As a function of the principle of complementarity, the completeness of the quantum formalism and the need to avert worries about the nonlocality of the collapse process, the wavefunction according to the Copenhagen interpretation can be thought of as simply ``a mathematical description of the state of observer knowledge'' \citep[p.~228]{Cramer88}.\footnote{Cramer's reading of the Copenhagen interpretation in this respect is somewhat contentious. The ``knowledge interpretation'' of the wavefunction is claimed to be an integral element of the Copenhagen interpretation by \citet{Heisenberg55} but this may be in conflict with the way Bohr envisaged the wavefunction. See \citet{Howard04} for an excellent discussion of this issue.\label{fn:heisbohr}} In contrast, the transactional interpretation takes the wavefunction to be a real physical wave with spatial extent.\footnote{Recent work by \citet{Kastner10} and \citet{KastnerCramer} suggests a potentially improved reading of the transactional interpretation where the wavefunction is considered as ``residing in a `higher' or external ontological realm corresponding to the Hilbert space of all quantum systems involved''. This is an interesting and potentially fruitful avenue for avoiding the problems I outline below concerning Cramer's realistic interpretation. I unfortunately do not take account of this improved reading here.} The wavefunction of the quantum mechanical formalism is identical with the initial offer wave of the transaction mechanism and the collapsed wavefunction is identical with the completed transaction. Quantum particles are thus not to be thought of as represented by the wavefunction but rather by the completed transaction, of which the wavefunction is only the initial phase. As Cramer explains:
\begin{quote}
  \small
  \singlespacing
  The transaction may involve a single emitter and absorber or multiple emitters and absorbers, but it is only complete when appropriate boundary conditions are satisfied at all loci of emission and absorption. Particles transferred have no separate identity independent from the satisfaction of these boundary conditions. \citeyearpar[p.~666]{Cramer}
\end{quote}
Though there is much formal overlap between particular elements of the Copenhagen and transactional interpretations, Cramer points out that giving objective reality to the wavefunction ``colors all the other elements of the interpretation'' leading to a vastly different physical picture of the quantum world. Let us consider this physical picture with a concrete example.

\begin{figure}[t]
  \begin{center}
    \begin{tikzpicture}
      \small
      \path[use as bounding box] (0,0) rectangle (14,6);
      \draw[->] (1,1.5) -- (1,3.5) node[above=2mm] {time};
      \filldraw (4,1) circle (0.3mm) node[below=2mm] {$B:(\mathbf{x}_{B},t_{0})$};
      \filldraw (7,1) circle (0.3mm) node[below=2mm] {$S:(\mathbf{x}_{0},t_{0})$};
      \filldraw (10,1) circle (0.3mm) node[below=2mm] {$A:(\mathbf{x}_{A},t_{0})$};
      \filldraw (4,5) circle (0.3mm) node[above=2mm] {$(\mathbf{x}_{B},t_{1})$};
      \filldraw (10,5) circle (0.3mm) node[above=2mm] {$(\mathbf{x}_{A},t_{1})$};
      \draw (8,1) arc (0:180:1);
      \draw (7,2) node[above] {$\psi_{o}(\mathbf{x},t)$};
      \draw (11,5) arc (360:180:1);
      \draw (10,4) node[below left] {$\psi_{c_{A}}(\mathbf{x},t)$};
      \draw (5,5) arc (360:180:1);
      \draw (4,4) node[below right] {$\psi_{c_{B}}(\mathbf{x},t)$};
    \end{tikzpicture}
  \end{center}
  \caption{Offer and confirmation waves in the transactional interpretation}
  \label{fig:cramexp}
\end{figure}
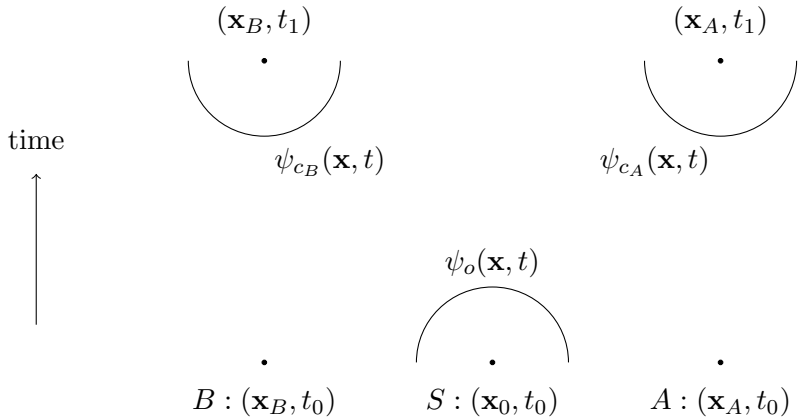

Consider a radioactive source, S, sitting between two absorbers, A and B, constrained to emit a single $\beta$-particle either to the left or to the right (Figure~\ref{fig:cramexp}). According to the transactional interpretation, the process of $\beta$-particle emission can be described in terms of offer and confirmation waves, the initial offer wave being the wavefunction of the quantum mechanical formalism. The wavefunction ``is a real physical wave generated by the emitter, and travels through space to the final absorber as well as to many other spacetime loci and many other potential absorbers'' \citeyearpar[p.~667]{Cramer}. Thus at the time of emission, $t_{0}$, an offer wave is produced which propagates towards each absorber as well as forwards and backwards in time. Upon being stimulated by this offer wave at time $t_{1}$, the absorbers A and B each produce a confirmation wave that propagates backwards in time (among other directions) to the radioactive source; the amplitude of each confirmation wave evaluated at the source is proportional to the modulus squared of the amplitude of the offer wave evaluated at the respective absorber (i.e. $|\psi_{c_{i}}(\mathbf{x}_{0},t_{0})|\propto|\psi_{o}(\mathbf{x}_{i},t_{1})|^{2}$ for $i=A,B$). These confirmation waves provide the emitter, so to speak, with a Born probability measure over which the likelihood of each particular transaction occurring can be quantified. In the same way that the absorber responds to the initial offer wave, the emitter responds to this subsequent confirmation wave and this cycle continues.

Let us say that the transaction is completed between the radioactive source and absorber A, i.e. a four dimensional standing wave emerges between S at $t_{0}$ and A at $t_{1}$. The components of the wavefunction which permeate the spatiotemporal regions that are not between S at $t_{0}$ and A at $t_{1}$ do not ``disappear''. Rather these components ``are only virtual in the sense that they transfer no energy or momentum and participate in no transaction''. Moreover, ``the emergence of this transaction does not occur at any particular location in space or at some particular instant in time, but rather forms along the entire four-vector that connects the emission locus with the absorption locus'' \citeyearpar[p.~667]{Cramer}. This four dimensional standing wave is then interpreted as the emission of a $\beta$-particle to the left by S at $t_{0}$ and the subsequent absorption of this $\beta$-particle by A at $t_{1}$.

The transactional interpretation of the quantum formalism allows the resolution of some of the most worrying aspects of the Copenhagen interpretation. Since we do not require the disappearance of the initial wavefunction upon completion of the transaction, the transactional interpretation alleviates the need to resort to an epistemic interpretation of the wavefunction, which \citet[p.~228]{Cramer88} finds ``intellectually unappealing'', to account for the nonlocality associated with wavefunction collapse.\footnote{See fn.~\ref{fn:heisbohr}.} In addition, the transactional interpretation subverts the dilemma at the core of the EPR argument (\citet{EPR}) by permitting the simultaneous reality of incompatible operators: the wavefunction, according to the transactional interpretation,
\begin{quote}
  \small
  \singlespacing
  brings to each potential absorber the full range of possible outcomes, and all have ``simultaneous reality'' in the EPR sense. The absorber interacts so as to cause one of these outcomes to emerge in the transaction, so that the collapsed [wavefunction] manifests only one of these outcomes. \citeyearpar[p.~668]{Cramer}.
\end{quote}
Most importantly, however, the transactional interpretation employs both retarded and advanced waves, and in doing so admits the possibility of providing a `zigzag' explanation of the nonlocality associated with entangled quantum systems. The boundary conditions that influence the formation of a completed transaction include both those at the emitter as well as the future absorbers. It is this feature that makes the transactional interpretation a retrocausal model of quantum mechanics. Moreover, it is this feature that enables the combination of two or more local influences to yield a nonlocal influence, which restores Lorentz covariance to the description of the behaviour of entangled quantum systems.

While it at least appears as though the transactional interpretation goes some way to resolving the interpretational issues of the Copenhagen interpretation, it is in fact not without its own points of contention.

\section{Maudlin's objection}
\label{sec:maudlin}

\citet{Maudlin} outlines a selection of problems that arise in Cramer's theory as a result of the pseudotemporal account of the transaction mechanism: processes important to the completion of a transaction take place in pseudotime only (rather than in real time) and thus cannot be said to have taken place at all. Since a temporally bound observer can only ever perceive a completed transaction, i.e. a collapsed wavefunction, the uncollapsed wavefunction never actually exists. Since the initial offer wave is identical to the wavefunction of the quantum formalism, any ensuing exchange of advanced and retarded waves required to provide the quantum mechanical probabilities, according to Maudlin, also do not exist. Moreover, Cramer's exposition of the transaction mechanism seems to suggest that the stimulation of sequential offer and confirmation waves occurs deterministically, leaving a gaping hole in any explanation the transactional interpretation might provide of the stochastic nature of quantum mechanics. Although these problems are significant, Maudlin admits that they may indeed be peculiar to Cramer's theory. Having said this, Maudlin also sets out a more general objection to retrocausal models of quantum mechanics which he claims to pose a problem for ``any theory in which both backwards and forwards influences conspire to shape events'' \citeyearpar[p.~201]{Maudlin}.

Maudlin's main objection to the transactional interpretation hinges upon the fact that the transaction process depends crucially on the fixity of the absorbers ``just sitting out there in the future, waiting to absorb'' \citeyearpar[p.~199]{Maudlin}; one cannot presume that present events are unable to influence the future disposition of the absorbers. Let us consider Maudlin's own thought experiment designed to illustrate this objection.
\begin{figure}[t]
  \begin{center}
    \begin{tikzpicture}
      \small
      \filldraw (7,5) circle (0.3mm) node[below=2mm] {S};
      \filldraw (10,5) circle (0.3mm) node[below=2mm] {A};
      \filldraw (13,5) circle (0.3mm) node[below=2mm] {B};
      \draw[->] (7.5,5) -- (8.5,5) node[right] {$\beta$};
      \draw (7,4) node[below] {\textbf{(i)}};
      \filldraw (7,2) circle (0.3mm) node[below=2mm] {S};
      \filldraw (10,2) circle (0.3mm) node[below=2mm] {A};
      \filldraw (1,2) circle (0.3mm) node[below=2mm] {B};
      \draw[->] (3.5,2) -- (2.5,2) node[left] {$\beta$};
      \draw[->] (13,2.2) .. controls (13,3) and (1,3) .. (1,2.2);
      \draw (7,1) node[below] {\textbf{(ii)}};
    \end{tikzpicture}
  \end{center}
  \caption{Maudlin's thought experiment}
  \label{fig:maudexp}
\end{figure}
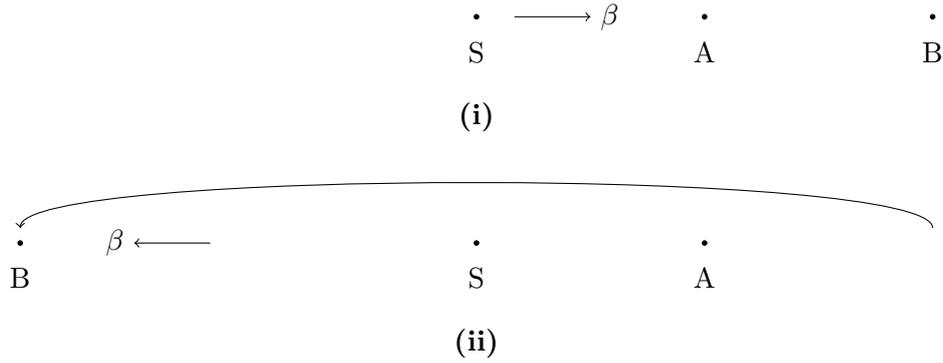
Consider again our radioactive source constrained to emit a $\beta$-particle either to the left or to the right. To the right sits absorber A at a distance of 1 unit. Absorber B is also located to the right but at a distance of 2 units and is built on pivots so that it can be swung around to the left on command (Figure~\ref{fig:maudexp}(i)). A $\beta$-particle emitted at time $t_{0}$ to the right will be absorbed by absorber A at time $t_{1}$. If after time $t_{1}$ the $\beta$-particle is not detected at absorber A, absorber B is quickly swung around to the left to detect the $\beta$-particle after time $2t_{1}$ (Figure~\ref{fig:maudexp}(ii)).

According to the transactional interpretation, since there are two possible outcomes (detection at absorber A or detection at absorber B), there will be two confirmation waves sent back from the future, one for each absorber. Furthermore, since it is equally probable that the $\beta$-particle be detected at either absorber, the amplitudes of these confirmation waves should be equal. However, a confirmation wave from absorber B can only be sent back to the emitter if absorber B is located on the left. For this to be the case, absorber A must not have detected the $\beta$-particle and thus the outcome of the experiment must already have been decided. The incidence of a confirmation wave from absorber B at the emitter \emph{ensures} that the $\beta$-particle is to be sent to the left, even though the amplitude of this wave implies a probability of a half of this being the case. As Maudlin states so succinctly, ``Cramer's theory collapses''.

It is clear to see that this challenge to retrocausality must be considered seriously if a proposed retrocausal mechanism is to be successful. The key challenge from Maudlin is that any retrocausal mechanism must ensure that the future behaviour of the system transpires consistently with the spatiotemporal structure dictated by any potential future causes: ``stochastic outcomes at a particular point in time may influence the future, but that future itself is supposed to play a role in producing the outcomes'' \citeyearpar[p.~197]{Maudlin}. In the transactional interpretation the existence of the confirmation wave itself presupposes some determined future state of the system with retrocausal influence. However, with standard (i.e. forwards-in-time) stochastic causal influences affecting the future from the present, a determined future \emph{may not necessarily be guaranteed} in every such case, as shown by Maudlin's experiment. Before we go on to examine this objection in more detail, let us first consider some responses that have been put forward in defence of Cramer's theory.

\section{Cramer defended}
\label{sec:defend}

I wish to examine here three specific defences of the transactional interpretation due to \citet{Berkovitz}, \citet{Kastner} and \citet{Marchildon}. A review of these defences will not only provide a good exercise in exploring the details of the transactional interpretation, but will assist us in getting to the source of the issues highlighted by Maudlin's challenge.

Maudlin's objection has been formulated by \citet{Berkovitz} in terms of the varying conceptualisations of the probabilities involved in the experiment. More specifically, Berkovitz believes that the deviation between the long-run frequencies of measurement outcomes and their objective probabilities is at the core of the objection. Berkovitz defends the transactional interpretation by showing that causal loops of the type found in Maudlin's experiment need not obey the assumptions about probabilities that are common in linear causal situations. To illustrate this claim about causal loops, Berkovitz considers a simple coin toss.

Let the event P be the tossing of a fair coin, and let this be an indeterministic cause of event Q, the coin landing `heads'. Let event R be the perception of the coin landing `heads' deterministically caused by event Q. Since the coin is fair, the long-run frequency of event Q with respect to event P is $\tfrac{1}{2}$. However, if one considers the long-run frequency of event Q with respect to both P \emph{and} R, then this frequency is 1; every time event P occurs with event R, Q must have occurred. The probability of event Q with respect to P and R is called by Berkovitz a biased probability. Berkovitz argues that within causal loops of the type found in Maudlin's experiment the probabilities are always biased. Thus one should not expect the long-run frequencies to correspond with any unbiased probabilities; there is no inconsistency in a deviation between these quantities.

This example can be translated in a straightforward manner to the language of Maudlin's experiment. Event P is the radioactive $\beta$-decay, event Q (indeterministically caused by P) is the emission of the $\beta$-particle to the left and event R (deterministically caused by Q) is the detection of the $\beta$-particle on the left. Recall that an integral element of Maudlin's objection is that the existence of the confirmation wave on the left \emph{ensures} event Q, but the information contained within the confirmation wave itself suggests event Q has a probability of only $\tfrac{1}{2}$. With respect to only event P, event Q has a long-run frequency of $\tfrac{1}{2}$, but with respect to both P and R this biased probability is 1. It is not inconsistent for these quantities to deviate, therefore Berkovitz claims Cramer's theory is not inconsistent.

Berkovitz does not consider Cramer's pseudotemporal account of the transaction mechanism significant, preferring to think of the cycle of offer and confirmation waves in terms of causal connections which are part of a four dimensional block universe. While Berkovitz has claimed to show the legitimacy of the causal loop in Maudlin's experiment, by overlooking Cramer's pseudotemporal account of the transaction mechanism Berkovitz has neglected to address exactly how the pseudotemporal account \emph{can} be consistent with the four dimensional block universe. Berkovitz returns to the transactional interpretation in his \citeyearpar{Berkovitz08} where he recognises that the pseudotemporal account of the transaction mechanism jeopardises the explanatory value of the theory. However, the ontological nature of the transaction mechanism is once again left to one side in his analysis.

\citet{Kastner} has expanded on Berkovitz' approach with a view to eliminating pseudotime from the transactional interpretation. Kastner begins by noting that in the transactional interpretation a \emph{complete} set of absorbers is not necessary; it is possible for no confirmation wave to be received from the left of the radioactive source in Maudlin's experiment. Kastner differentiates between the initial states of the radioactive source in the two situations where (i) a confirmation wave is received from both the right and the left absorbers (absorbers A and B respectively), and (ii) a confirmation wave is received from only the right absorber (absorber A only). It is clear that if a confirmation wave is received from both the left and right then it is the case that the $\beta$-particle will be emitted to the left. Recall that Maudlin claims this to be inconsistent with the information contained in the confirmation wave from the left.

In a similar fashion to the analysis of Berkovitz, Kastner emphasises the disparity of probabilities as the heart of Maudlin's objection to the transactional interpretation. However, given the initial state, according to Kastner, this disparity can be explained. The probability of emission to the left in the case where a confirmation wave is received from both the left and the right is $\tfrac{1}{2}$ according to the information contained in each confirmation wave. However, the probability of this being the initial state of the emitter is also $\tfrac{1}{2}$ since there are two equally probable initial states. Thus, using the standard probabilistic expression, the probability of emission to the left given the initial state is
\begin{equation}
  \nonumber
  \frac{P(L\mathbin{\&}\psi)}{P(\psi)} = \frac{\frac{1}{2}}{\frac{1}{2}} = 1,
\end{equation}
where L is emission to the left and $\psi$ is the initial state.

These two initial states of Maudlin's experiment can be imagined, according to Kastner, as belonging to two distinct worlds, which share only the offer and confirmation waves between the emitter and absorber A in common. Kastner proposes that the incipient transaction corresponding to the offer and confirmation waves between the emitter and absorber A can be thought of as an unstable bifurcation line between the two worlds. The success or failure of this transaction determines which world the system ``enters''. Suppose the incipient transaction between the emitter and absorber A fails. If this is the case, absorber A does not detect the $\beta$-particle and absorber B is swung around to the left where it is now able to emit a confirmation wave. What would otherwise have been a null outcome becomes a realised transaction.

Kastner points out that this account must ``abandon the idea that there is cyclic `echoing' between absorber B and the emitter if such echoing is taken as reflective of an uncertainty in outcome'' \citeyearpar[p.~14]{Kastner}. The failure of the bifurcating transaction, i.e. that between the emitter and absorber A, makes the outcome of emission to the left certain. Moreover, the information contained in the confirmation wave received from absorber B indicates a probability of a half of this being the case. This is not inconsistent due to the above analysis and, in fact, shows that each confirmation wave reflects the probability structure across both possible worlds, demonstrating the holistic structure of quantum mechanics.

\citet{Marchildon} proposes another defence of the transactional interpretation against Maudlin's objection. He begins by supposing another absorber, say C, is situated on the left of the radioactive source in Maudlin's thought experiment at a distance larger than that of absorber B from the source. If this is the case, then the emitter will receive a confirmation wave from absorber C on the left and Maudlin's experiment will proceed as usual. Marchildon then proposes removing absorber C and considering the absorption properties of the long distance boundary conditions. If it is postulated that the universe is a perfect absorber of all radiation then the presence of absorber C is irrelevant; a confirmation wave from the left will always be received by the radioactive source at the time of emission and it will encode the correct probabilistic information. This enables the transactional interpretation to remain consistent in Maudlin's experiment. On the assumption that the universe is a perfect absorber, the transactional interpretation correctly predicts that the $\beta$-particle will be emitted to the left half of the time. It remains the case, however, that the transaction is completed with absorber B only if it is situated on the left. According to Marchildon, ``although the confirmation wave coming from the left originates from the remote absorber just as often as it originates from B, the transaction is never completed with the remote absorber'' \citeyearpar[p.~12]{Marchildon}.

Although there is a varied focus to each of these defences of Cramer, it is clear that the problematic element of the transactional interpretation is the causal structure of the pseudotemporal account of the transaction mechanism. In the next section I offer an analysis of Maudlin's experiment according to this pseudotemporal account from the perspective of the block universe model. In doing so I hope to show why Maudlin's experiment still poses a problem for Cramer's theory despite these defences. The underlying problem is that Cramer's theory fails to provide a sufficient causal structure to constrain uniquely the behaviour of the system. While I think Maudlin has successfully isolated this shortcoming, in \S\ref{sec:symmetry} I challenge his justification for why this is the case in the transactional interpretation.

\section{Maudlin's experiment in four dimensions}
\label{sec:fourdim}

The central claim with which Berkovitz and Kastner are concerned is the disparity between the probability of emission to the left as determined by the amplitude of the confirmation wave from the left absorber and the expected probability given that a confirmation wave arrives from the left. Above I characterised Maudlin's objection in a different manner: Maudlin's key challenge is that any retrocausal mechanism must ensure that the future behaviour of the system transpires consistently with the spatiotemporal structure dictated by any potential future causes. An instructive way to analyse the causal structure of Maudlin's experiment is four dimensionally.

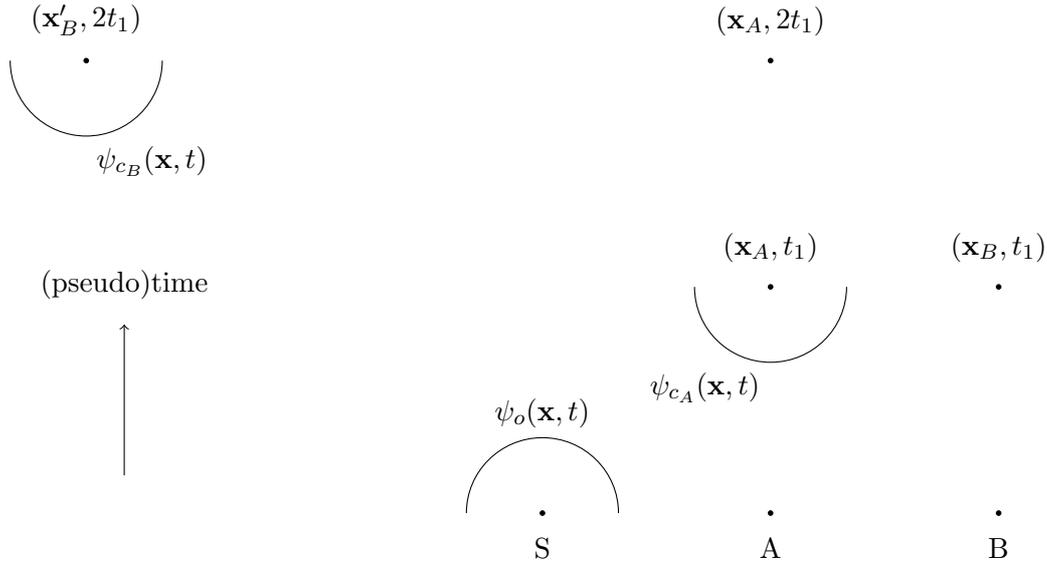
\begin{figure}[t]
  \begin{center}
    \begin{tikzpicture}
      \small
      \path[use as bounding box] (0,0) rectangle (14,8);
      \draw[->] (1.5,1.5) -- (1.5,3.5) node[above=2mm] {(pseudo)time};
      \filldraw (7,1) circle (0.3mm) node[below=2mm] {S};
      \filldraw (10,1) circle (0.3mm) node[below=2mm] {A};
      \filldraw (13,1) circle (0.3mm) node[below=2mm] {B};
      \filldraw (10,4) circle (0.3mm) node[above=2mm] {$(\mathbf{x}_{A},t_{1})$};
      \filldraw (13,4) circle (0.3mm) node[above=2mm] {$(\mathbf{x}_{B},t_{1})$};
      \filldraw (10,7) circle (0.3mm) node[above=2mm] {$(\mathbf{x}_{A},2t_{1})$};
      \filldraw (1,7) circle (0.3mm) node[above=2mm] {$(\mathbf{x}^{\prime}_{B},2t_{1})$};
      \draw (8,1) arc (0:180:1);
      \draw (7,2) node[above] {$\psi_{o}(\mathbf{x},t)$};
      \draw (11,4) arc (360:180:1);
      \draw (10,3) node[below left] {$\psi_{c_{A}}(\mathbf{x},t)$};
      \draw (2,7) arc (360:180:1);
      \draw (1,6) node[below right] {$\psi_{c_{B}}(\mathbf{x},t)$};
    \end{tikzpicture}
  \end{center}
  \caption{Maudlin's thought experiment according to the transactional interpretation}
  \label{fig:block}
\end{figure}

Consider once again Maudlin's experimental setup and let us imagine a $\beta$-particle emission to the right (i.e. towards absorber A) according to the transactional interpretation. Figure~\ref{fig:block} represents that part of the transaction process that occurs in pseudotime (a sort of `space-pseudotime' diagram). An offer wave is emitted from the radioactive source at time $t_{0}$. If we initially ignore the conditional nature of the event structure of the experiment, we can imagine this offer wave stimulating two confirmation waves from absorbers A and B, each confirmation wave originating from the respective potential absorber positions in spacetime. The particular transaction process we are considering determines that a four dimensional standing wave emerge between absorber A at time $t_{1}$ and the source S at time $t_{0}$, which is interpreted as the emission of a $\beta$-particle at S and the absorption of this particle at A. The passage of the $\beta$-particle, whose four-vector emerges atemporally over the entire locus of the transaction, is a process of spacetime while the transaction mechanism itself is a process of pseudotime. If we now consider the conditional nature of the event structure in time, due to the absorption event at absorber A, absorber B will remain on the right. Curiously the four dimensional `space-pseudotime' block contains an event structure (i.e. absorber B swinging to the left) which the four dimensional spacetime block does not. Both absorber A and absorber B vie \emph{in pseudotime} to participate in the completed transaction but, once the transaction emerges for one absorber only, the standing wave that is formed is a standing wave \emph{in spacetime}. We should acknowledge at this point that Cramer attempts to alleviate such a worry by suggesting pseudotime to be ``a pedagogical convention for the purposes of description'' (see \S\ref{sec:transaction}).

However, this worry is nonetheless compounded by Cramer's insistence that the initial offer wave, identical to the wavefunction of the quantum formalism, is a real wave that propagates through space. Recall that the components of this offer wave that do not participate in any eventual transaction are described by Cramer as ``virtual'' in that they transfer no energy or momentum. What Cramer fails to account for is the fact that these virtual components do contribute something quite important to the transaction mechanism: a putative causal structure. The role that is played by those components of the wavefunction that are not emitted in the direction of the eventual absorber is to stimulate virtual confirmation waves which in turn provide the emitter with the relevant Born probability measure over all future possibilities. This potential causal influence, however, originates from a `space-pseudotime' event structure that is not necessarily representative of the event structure in the future of the quantum system; the probability measure is constrained by objects that may not physically be there! There is then something very strange in claiming that the virtual cycles of offer and confirmation waves play a causal role in constraining the event structure of spacetime. There seems to be a mismatch between the causal structure dictated by the initial conditions and the causal structure dictated by the actual evolution of the system which turns on the obscure ontological status of the pseudotemporal process.

Maudlin's objection is a potent one; the pseudotemporal account of the transaction mechanism in the transactional interpretation resists straightforward clarification. This somewhat complicates the defences examined in the last section, particularly those of Berkovitz and Marchildon. Berkovitz does consider the cycle of offer and confirmation waves between emitters and absorbers in the context of a four dimensional block universe \citeyearpar[p.~242]{Berkovitz}, but does so without considering the reality of these entities within the spacetime block. By playing an important role in the transaction mechanism, the offer and confirmation waves have causal significance in Berkovitz' causal loops. However, it is this causal significance that is called into question by Maudlin's objection rather than the consistency of the causal loops.

In contrast, Marchildon eschews any causal significance of the pseudotemporal offer and confirmation waves by assuming the universe to be a perfect absorber; there will always be \emph{some} confirmation wave returning to the source at the time of emission from every direction, which can play the role of providing the Born probability measure. However, this does not do justice to the pseudotemporal account of the transaction mechanism. Recall that ``the emergence of [the completed] transaction does not occur at any particular location in space or at some particular instant in time, but rather forms along the entire four-vector that connects the emission locus with the absorption locus''. Thus the emergence of the completed transaction \emph{just is} the emission of the $\beta$-particle, the passage of the $\beta$-particle from emitter to absorber and then the absorption of the $\beta$-particle, all together. The emergence of the transaction is \emph{not} the emission event. Therefore the perfectly absorbing universe cannot stand in for absorber B on the left because spacetime only contains completed transactions and completed transactions are, by construction, complete four-vector particle trajectories.

Perhaps Kastner is on the right track by attempting to eliminate pseudotime from the interpretation by eliminating the dependence of the transaction mechanism on the position of \emph{all} the possible absorbers. The resulting view of bifurcating worlds, however, is metaphysically rather strange. Indeed, in some sense there may be a correspondence between Kastner's portrayal of bifurcating worlds and the above adoption of `space-pseudotime' diagrams. There certainly seems to be a need to account for a multitude of event structures precipitated by the pseudotemporal account of the transaction mechanism. I intimated above that this counterfactual feature of the transaction mechanism is at odds with Cramer's insistence on a real wavefunction. One might argue, if one was that way inclined metaphysically, that these facets of the interpretation can be made coherent by allowing for bifurcating worlds such as Kastner's.

Unless one were that way inclined, however, Maudlin's objection remains damaging. I do not, though, consequently follow Maudlin in thinking that ``any theory in which both backwards and forwards influences conspire to shape events will face this same challenge''. Recall that the selection of problems introduced in \S\ref{sec:maudlin} that Maudlin thought peculiar to Cramer's theory arose as a result of the pseudotemporal account of the transaction mechanism. These problems are intimately linked, I think: the pseudotime heuristic and the reality of the wavefunction are difficult to reconcile. As we have just seen, however, a more significant worry is the underconstrained nature of the behaviour of the system. Maudlin believes this to be endemic to retrocausal theories in general. I contend that it is the lack of causal symmetry in Cramer's theory that is to blame here.

\section{Causal symmetry}
\label{sec:symmetry}

The pseudotemporal account of the transaction mechanism that Cramer provides, while retrocausal in the sense that it contains both retarded and advanced influences, is not time symmetric. The initial offer wave always precedes (pseudotemporally) the other processes of the transaction and thus the initial conditions of a quantum system described by the transactional interpretation have primacy over any other boundary condition in constraining the dynamics. This is instrumental in rendering the pseudotemporal account of the transaction mechanism problematic. The varying event structures associated with different possible outcomes of a single stochastic event that we encountered above would not arise if the transaction mechanism endowed both the retarded and advanced elements of the transaction with equivalent causal significance. To do so would amount to constraining the transaction mechanism from both temporal ends and this, in turn, would be enough to constrain the event structure uniquely in spacetime.

Indeed, Maudlin suggests something along these lines as the key to a successful retrocausal theory:
\begin{quote}
  \small
  \singlespace
  If the course of present events depend on the future and the shape of the future is in part determined by the present then there must be some structure which guarantees the existence of a coherent mutual adjustment of all the free variables. \citeyearpar[p.~201]{Maudlin}
\end{quote}
Thus due to the causal asymmetry of the pseudotemporal account of the transaction mechanism, the retarded and advanced elements of Cramer's theory demonstrably do \emph{not} have a structure which guarantees the existence of a coherent mutual adjustment of all the free variables.

Maudlin realises that this failure to provide a coherent mutual adjustment of free variables is indeed the cardinal problem of the transactional interpretation, but suggests that the reason for this is simply because it is retrocausal. According to Maudlin, in theories without retrocausation (which Maudlin, following Bell, calls `local' theories),
\begin{quote}
  \small
  \singlespace
  solutions to the field equations at a point are constrained only by the values of quantities in one light cone (either past or future) of a point. Thus in a deterministic theory, specifying data along a hyperplane of simultaneity suffices to fix a unique solution at all times, past and future of the plane. Further, the solutions can be generated sequentially: the solution at $t=0$ can be continued to a solution at $t=1$ without having had to solve for any value at times beyond $t=1$. Thus the physical state at one time generates states at all succeeding times in turn\dots

  [In a stochastic theory] fixing the physical state in the back light cone of a point may not determine the physical state there, but it does determine a unique probability measure over the possible states such that events at spacelike separation are statistically independent of one another\dots The present moment makes all of its random choices independently and then generates the probabilities for the immediate future, and so on. \citeyearpar[p.~201]{Maudlin}
\end{quote}
He continues,
\begin{quote}
  \small
  \singlespace
  Any theory with both backwards and forwards causation cannot have such a structure. Data along a single hypersurface do not suffice to fix the immediate future since that in turn may be affected by its own future. The metaphysical picture of the past generating the future must be abandoned, and along with it the mathematical tractability of local theories. \citeyearpar[p.~201]{Maudlin}
\end{quote}

Maudlin's argument against retrocausality can thus be construed as follows:
\begin{enumerate}
  \item retrocausal theories must have a structure which guarantees the coherent mutual adjustment of free variables;
  \item local theories are mathematically tractable and fit a metaphysical picture of the past generating the future because solutions to the field equations require only data along a single hypersurface;
  \item data along a single hypersurface are not sufficient for a retrocausal theory to guarantee the coherent mutual adjustment of free variables;
  \item therefore, retrocausal theories must abandon both the metaphysical picture of the past generating the future and with it mathematical tractability.
\end{enumerate}

There are multiple reasons to be wary of this argument, which we will address here in turn. As a starting point, let us consider the claim that retrocausal theories must have a structure which guarantees the coherent mutual adjustment of free variables and that, because of this, retrocausal theories will be underdetermined by the data along a single hypersurface. Insofar as this is the case, the transactional interpretation can be seen as an attempt to achieve the former but with a failed mechanism for remedying the latter. The failure of the transactional interpretation to achieve this, however, is not because it is a retrocausal theory, rather, as indicated above, it is because it lacks causal symmetry in its pseudotemporal account of the fundamental quantum causal mechanism. I made the suggestion above that temporal symmetry could be achieved by the transactional interpretation if both initial and final boundary constraints were employed. It should now be clear that if such constraints were present in the formalism of a retrocausal theory then this would also debase any underdetermination claim; an increase in the available data would suffice to determine uniquely the behaviour of the system. It is not the case that such a retrocausal theory would, despite Maudlin's declaration to the contrary, elicit the abandonment of mathematical tractability; we will explore this further in just a moment. A related concern, however, is that it is not entirely clear that Maudlin's underdetermination claim should worry us in the first place.

\subsection{Causality and determination}

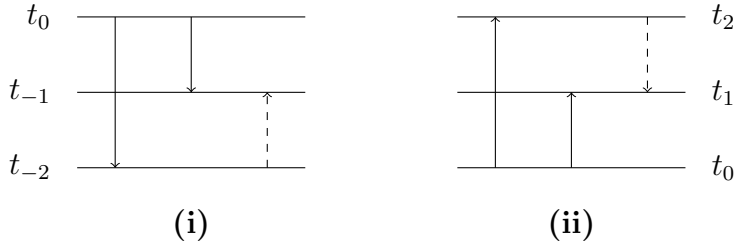
\begin{figure}[t]
  \begin{center}
    \begin{tikzpicture}
      \draw (4,3) -- (1,3) node[left=2mm]{$t_{0}$};
      \draw (4,2) -- (1,2) node[left=2mm]{$t_{-1}$};
      \draw (4,1) -- (1,1) node[left=2mm]{$t_{-2}$};
      \draw[->] (1.5,3) -- (1.5,1);
      \draw[->] (2.5,3) -- (2.5,2);
      \draw[dashed,->] (3.5,1) -- (3.5,2);
      \draw (2.5,1) node[below=4mm] {\textbf{(i)}};
      \draw (6,3) -- (9,3) node[right=2mm]{$t_{2}$};
      \draw (6,2) -- (9,2) node[right=2mm]{$t_{1}$};
      \draw (6,1) -- (9,1) node[right=2mm]{$t_{0}$};
      \draw[->] (6.5,1) -- (6.5,3);
      \draw[->] (7.5,1) -- (7.5,2);
      \draw[dashed,->] (8.5,3) -- (8.5,2);
      \draw (7.5,1) node[below=4mm] {\textbf{(ii)}};
    \end{tikzpicture}
  \end{center}
  \caption{Determination and causation in (i) ordinary forwards-in-time causal theories and (ii) retrocausal theories. The black arrows indicate determination and the dashed arrows indicate what we would like to think of as causal influences in those cases.}
  \label{fig:determine}
\end{figure}

Consider Maudlin's reasoning that, in a retrocausal setting, data along a single hypersurface do not suffice to fix the immediate future since that in turn may be affected by its own future. If such a feature of retrocausal theories cannot guarantee the coherent mutual adjustment of free variables, then by symmetry so should the temporal reverse of this reasoning fail to guarantee the coherent mutual adjustment of free variables in ordinary forwards-in-time causal cases, i.e. data along a single hypersurface should not suffice to fix the immediate past since that in turn may be affected by its own past.\footnote{See also \citet*{EvansPriceWharton} for the same point.} This is clearly not correct. In a deterministic theory, data along a single hypersurface are sufficient to determine a unique solution to the field equations and thus determine the behaviour of the system at all times, past and future. Thus the data at some time $t_{0}$ determine not only the data at $t_{-1}$ but also the data at $t_{-2}$, which is normally thought to have a causal influence on the data at $t_{-1}$ (see Figure~\ref{fig:determine}(i)). We see quite clearly here that the data at $t_{-2}$ is not an independent condition of the sort that could stymie the coherent mutual adjustment of free variables. By the same token the data at $t_{0}$ determine not only the data at $t_{1}$ but also the data at $t_{2}$, which in a retrocausal setting can be thought to have a causal influence on the data at $t_{1}$ (see Figure~\ref{fig:determine}(ii)). We can now see just as clearly that the data at $t_{2}$ is likewise not an independent condition of the sort that Maudlin claims renders retrocausal theories underdetermined.

It appears as though Maudlin's mistaken underdetermination claim emerges from a manifest tension between the temporal asymmetry of his ``metaphysical picture of the past generating the future'' and the temporal symmetry of determination in which ``data along a hyperplane of simultaneity suffices to fix a unique solution at all times, past and future of the plane''. The tension stems from the distinctly causal notion of ``generation'' in Maudlin's metaphysical picture in contrast to the ``fixity'' of a unique solution in his characterisation of determinism. We can alleviate this tension with a more carefully constructed picture of reality.\footnote{See \citet{PriceWeslake} and \citet{Evans} for a more detailed exposition of this picture.} If we characterise causality as a perspectival notion built upon an interventionist account of causation, we find that we are able to strike a harmony between our causal intuitions, such as deliberation, and the intuition that future events are fixed within a deterministic framework by realising that we, as spatiotemporally bound agents, are constrained in our epistemic access to the events in spacetime. The key to this picture is that our limited epistemic access to both the effects of our actions, on the one hand, and the complete causal past of our deliberations, on the other, allow us to believe that our `free' actions play a role in bringing about (or \emph{causing}) the consequences of our actions. With such a picture in mind, we are able to attribute both $t_{-2}$ and $t_{2}$ with causal significance insofar as we are ignorant of the complete data at $t_{-1}$ and $t_{1}$ respectively and at $t_{0}$. If we then utilise this surrogate picture to reconcile causality and determinism we can see by the above reasoning that Maudlin's argument for underdetermination loses some of its authority.

\subsection{A tractable alternative}

The details of this surrogate picture also arise in the context of another of Maudlin's claims. Let us for argument's sake grant that the underdetermination problem of retrocausal theories must be remedied and return to Maudlin's argument that doing so renders these theories mathematically intractable. Consider the Schr\"{o}dinger equation, the wave equation of nonrelativistic quantum mechanics: it is an example of an equation that requires only data along a single hypersurface to fix a unique solution. This is because it is first-order in time. If we consider the Klein-Gordon equation, which is second-order in time, we see that it requires twice the initial data to determine a solution. Ordinarily solutions to this classical scalar field equation are found by imposing two independent initial boundary conditions: the solution to the field equations at some particular time as well as the first time derivative of this solution. According to \citet{Wharton10}, the Klein-Gordon equation has resisted interpretation as a relativistic quantum mechanical wave equation partly due to this increase in required initial data. \citeauthor{Wharton10} makes the suggestion that a time symmetric approach to quantum mechanics can provide a natural resolution to this interpretational obstacle. Rather than imposing two independent initial boundary conditions on solutions to the Klein-Gordon equation, one can impose two boundary conditions \emph{at two different times}. This can be interpreted as supplying the field equations with data along two different instantaneous hypersurfaces or, likewise, as determining the behaviour of any system described by the Klein-Gordon equation with initial and final constraints. This causal symmetry, of course, is just the suggestion made above for overcoming Maudlin's underdetermination claim targeting retrocausal theories.\footnote{As well as \citet{Wharton10}, see also \citet{Sutherland} for an example of a retrocausal theory with a symmetric causal structure.}

Within Wharton's time symmetric scheme the full solution to the field equations cannot be known before any final constraint becomes epistemically accessible. The initial and final boundary conditions can be pictured as representative of consecutive external measurements on some quantum system. Without knowledge of the later measurement to be performed on the system one cannot solve the field equations and thus one cannot know the exact state of the system between the measurements. What one can know before the later measurement, however, is some best approximation to the full solution based on the initial data and it seems reasonable to think this would be the ordinary wavefunction of the Schr\"{o}dinger equation. This then yields a `hidden variable' theory of sorts where the wavefunction of the quantum formalism is interpreted as representing an observer's knowledge of the system and the full solution to the Klein-Gordon equation is interpreted as representing the actual state of the system, hidden from the observer. Upon measuring the system the observer gains knowledge of the final constraint and can retrodict the intervening state based on the now attainable full solution to the Klein-Gordon equation. This careful attention to the epistemic limitations of the spatiotemporally bound observer is just the same principle that buttresses our surrogate picture from above.

Moreover, such a retrocausal scheme for modelling quantum processes seems by no means `mathematically intractable'; on the contrary, not only do we have a straightforward algorithm for calculating the properties of any particular quantum system but we also have a clear metaphysical prescription, whose limitations reflect our limitations as spatiotemporally bound observers, for representing this system. We have seen that the metaphysical picture that Maudlin ties to mathematical tractability, that of the past generating the future, is abandoned trivially within any retrocausal scheme but this evidently does not imply that we must also abandon mathematical tractability. Indeed, by emphasising this traditionalist metaphysical picture of reality, it seems as though what Maudlin has in mind when he says `mathematical tractability' is something commensurate with a particular form of initial value problem. Retrocausal theories of quantum mechanics aside, if we look toward some of our more established physical theories we see that there is little justification for this characterisation of mathematical tractability.

\subsection{Classical tractability}

In the first place, the representation of the dynamical behaviour of classical physical systems according to analytical mechanics certainly does not preclude all but an initial value metaphysics. Granted, the Hamiltonian formulation of dynamics appears to provide good support for this metaphysical picture: the dynamical arena of Hamiltonian mechanics, phase space, is a space of possible initial values with a geometric structure that allows the determination of a unique dynamical path given any single point in the space. However, when one considers how this geometric structure is derived from the formalism of analytical mechanics, one finds that this Hamiltonian picture is merely (as \citet{Lanczos} points out) a ``remarkable simplification'' of a deeper dynamical picture. The geometric structure of phase space is encoded in Hamilton's equations of motion and, according to Lanczos, there are two ways that these equations can be derived. The first way is to decompose the second-order Lagrangian equations of motion into two first-order equations that can be transformed into Hamilton's equations by application of a Legendre transformation. The Lagrangian equations themselves are attained by way of the variational (or action) principle: the Lagrangian equations are the necessary and sufficient conditions for the action integral to remain stationary under arbitrary variations of the configuration of the system \emph{given the initial and final configurations of the system}. Thus it would seem that the Hamiltonian formulation might indeed be built upon temporally symmetric boundary conditions.

The second way of deriving Hamilton's equations of motion, however, observes that since the Legendre transformations are completely symmetric there is no requirement that we must take the Lagrangian formulation of mechanics as primary. As such, one can formulate Hamilton's equations directly without the Lagrangian equations nor the Legendre transformations \citep[p.~169]{Lanczos}. However, to do so one must produce a new action integral in terms of an extended set of independent variables and subject it once again to a variational principle; Hamilton's equations become the conditions for a stationary action integral under arbitrary variations which are again constrained by \emph{initial and final boundary conditions}. Regardless then of how one constructs the geometry of Hamiltonian phase space, the fundamental element of analytical mechanics remains the specification of initial and final boundary conditions as part of the variational principle. Thus it appears as unlikely that one might find justification for Maudlin's characterisation of `mathematical tractability' in analytical mechanics.

The case of general relativity is not so clear cut. On the one hand, it is more than reasonable to take the central lesson of general relativity to be that the fundamental ontological unit of our reality is a four dimensional solution to Einstein's field equations; solutions are clearly not obtained in Maudlin's `mathematically tractable' way.\footnote{See \citet[\S9.2.2]{Brown} for a discussion of this point.} On the other hand, though, considerable effort has been spent over the last half a century attempting to cast general relativity in a form that explicitly separates out a single temporal dimension from three spatial dimensions\footnote{See, for instance, \citet{Dirac58}, \citet{Bergmann}, \citet{Arnowitt} and \citet{BarbourI}.}, which would appear the best hope for a justification of Maudlin's metaphysical picture: as long as a spacetime is globally hyperbolic, a solution can be generated from data on any Cauchy surface. However, a new difficulty arises herein: due to the foliation invariance of such formulations of general relativity there exists a troublesome indeterminacy problem. Not only does specification of a single 3-geometry (of which a phase space point is comprised when combined with the relevant canonically conjugate momentum variable) fail to determine uniquely a dynamical path but, as \citet{Pooley} points out, ``the specification of an \emph{initial sequence} of 3-geometries is not sufficient to allow us to predict which continuation of the sequence will be actualized'' (emphasis added). This indeterminacy may not be as pernicious as it first appears, since it is a function of gauge freedom, and thus every actualised sequence represents the same spacetime sliced in different ways. However, at the level of hypersurfaces, data along a single hypersurface is insufficient to determine a unique continuation in its immediate future.\footnote{This is related to the thin and thick sandwich problems; see \citet*{BSW} and \citet{Wheeler}.} It is up for grabs then whether or not general relativity fits Maudlin's characterisation of `mathematical tractability'.

\section{Cramer's missing structure and Maudlin's misdirected metaphysics}

Maudlin's inventive thought experiment exposes a deep problem within Cramer's theory: the causal structure of the transaction mechanism cannot constrain uniquely and consistently the behaviour of particular quantum systems. I claim that what the transactional interpretation is missing is a causally symmetric account of the transaction mechanism: that is, both initial and final boundary constraints with equal causal significance influencing the dynamics of the system. Such a causally symmetric mechanism would serve to ensure the coherent mutual adjustment of all the relevant free variables. In contrast, Maudlin attributes this shortcoming of the transactional interpretation to the inability of a retrocausal theory to supply a structure that could achieve such mutual adjustment of variables. Moreover, Maudlin claims that the inability of retrocausal theories to achieve this is due to a fundamental incongruence between retrocausality and his ``metaphysical picture of the past generating the future''. This picture underpins his underdetermination challenge to retrocausal theories and his notion of mathematical tractability.

I hope to have shown that we have good reason to be wary of Maudlin's metaphysical picture and its connection to mathematical tractability. Firstly, Maudlin's underdetermination challenge to retrocausality can be subverted if one is careful to spell out the metaphysical difference between causality and determination; Maudlin's picture evidently does not achieve this. Secondly, we saw an example of a retrocausal theory of quantum mechanics that does not encounter any problems with mathematical tractability, despite not adhering to the edict of Maudlin's picture. Thirdly, it seems unlikely that analytical mechanics, and possibly general relativity, can be used to support an initial value metaphysics, despite being arguably the best place to begin looking for mathematical tractability in physical theories. At the very least we can conclude from this that Maudlin's picture cannot be used as strongly as he may have liked as an objection against retrocausality.

The transactional interpretation, and Maudlin's critique, does show us something important: for retrocausality to be taken seriously in contemporary physics, it must be supported by a coherent picture of reality and, above all, this picture would do well to be causally symmetric.

\end{document}